# Unconventional quantum vortex matter state hosts quantum oscillations in the underdoped high-temperature cuprate superconductors

Yu-Te Hsu,[1][*] Máté Hartstein,[1] Alexander J. Davies,[1]
Alexander J. Hickey,[1] Mun K. Chan,[2] Juan Porras,[3] Toshinao Loew,[3]
Sofia V. Taylor,[1] Hsu Liu,[1] Alexander G. Eaton,[1] Matthieu Le Tacon,[3,4]
Huakun Zuo,[5] Jinhua Wang,[5] Zengwei Zhu,[5] Gilbert G. Lonzarich,[1]
Bernhard Keimer,[3] Neil Harrison,[2][†] Suchitra E. Sebastian[1][†]

[1] Cavendish Laboratory, Cambridge University, J. J. Thomson Avenue, Cambridge CB3 0HE, UK

[2] Pulsed Field Facility, Los Alamos National Laboratory, Los Alamos,
Mail Stop E536, Los Alamos, NM 87545, USA

[3] Department of Solid State Spectroscopy, Max Planck Institute for Solid State Research,
Heisenbergstr. 1, D-70569 Stuttgart, Germany

[4] Institute for Quantum Materials and Technologies, Karlsruhe Institute of Technology,
Hermann-v.-Helmholtz-Platz 1, D-76344 Eggenstein-Leopoldshafen, Germany

[5] Wuhan National High Magnetic Field Center and School of Physics, Huazhong University
of Science and Technology, Wuhan 430074, China

---

[*] current address: High Field Magnet Laboratory (HFML-EMFL) and Institute for Molecules and Materials, Radboud University, Toernooiveld 7, 6525 ED Nijmegen, Netherlands

[†] To whom correspondence should be addressed. E-mail: nharrison@lanl.gov; suchitra@phy.cam.ac.uk

**A central question in the underdoped cuprates pertains to the nature of the pseudogap ground state. A conventional metallic ground state of the pseudogap region has been argued to host quantum oscillations upon destruction of the superconducting order parameter by modest magnetic fields. Here we use low applied measurement currents and millikelvin temperatures on ultra-pure single crystals of underdoped $YBa_2Cu_3O_{6+x}$ to unearth an unconventional quantum vortex matter ground state characterized by vanishing electrical resistivity, magnetic hysteresis, and non-ohmic electrical transport characteristics beyond the highest laboratory accessible static fields. A new model of the pseudogap ground state is now required to explain quantum oscillations that are hosted by the bulk quantum vortex matter state without experiencing sizeable additional damping in the presence of a large maximum superconducting gap; possibilities include a pair density wave.**

Previous electrical transport measurements in underdoped $YBa_2Cu_3O_{6+x}$[1] reported the occurrence of quantum oscillations above modest magnetic fields ≈ 22 T in a conventional metallic pseudogap ground state characterised by finite electrical resistivity, in which superconductivity is destroyed. Here we access the ground state of a new generation of pristine single crystals of underdoped $YBa_2Cu_3O_{6+x}$ (hole doping concentration $p = 0.108 - 0.132$) up to the highest accessible static magnetic field of 45 T and down to 40 mK in the limit of low applied measurement current density ($j \to 0$). We find instead the persistence of vanishing electrical resistivity characteristic of magnetic field-resilient superconductivity beyond the highest accessible static magnetic field of 45 T.

In this work, we measure pristine single crystals of underdoped $YBa_2Cu_3O_{6+x}$ ($YBCO_{6+x}$) characterised by prominent quantum oscillations with a significantly lower Dingle (impurity damping) factor than the previous generation of single crystals[2] (Fig. S1). We access the longitudinal voltage in the ground state by applying a range of applied measurement current densities $j$ down to low values of $\approx 0.01$ Acm$^{-2}$, three orders of magnitude lower than previously used values of $\approx 10$ Acm$^{-2}$. Fig. 1$A$-$H$ shows the effective resistivity ($\rho_{xx} = V_{xx}/(jl)$ where $l$ is the distance between the voltage leads both in the case of ohmic and non-ohmic transport characteristics) as a function of magnetic field and temperature. For sufficiently low applied measurement currents and temperatures, we find vanishing electrical resistivity that persists beyond the highest accessible static magnetic field of 45 T (Fig. 1$A$-$H$). A steeply rising phase boundary of the resistive magnetic field scale up to which vanishing electrical resistivity persists in the limit of low applied measurement current (i.e. $\mu_0 H_\mathrm{r}(j \rightarrow 0)$) as a function of temperature is shown in Fig. 2. A similarly sharply rising resistive magnetic field that shows no sign of saturation down to temperatures $\leq 0.001 T_\mathrm{c}(0)$ was reported by Mackenzie et al. in pristine single crystals of Tl$_2$Ba$_2$CuO$_{6+\delta}$ (Fig. S2),[3] which also exhibit quantum oscillations.[4]

The low values of critical current as a function of $\mu_0 H$ and $T$ that we find to characterise high magnetic field-resilient superconductivity in high purity single crystals of underdoped YBa$_2$Cu$_3$O$_{6+x}$ is consistent with a low density of pinning centres. Evidence of bulk superconductivity in the ground state is further revealed by magnetic hysteresis due to bulk vortex pinning in magnetic torque. We find good agreement between the resistive magnetic field in the limit of low applied measurement current (i.e. $\mu_0 H_\mathrm{r}(j \rightarrow 0)$), and the high magnetic field scale up to which hysteresis from bulk vortex pinning persists in the magnetic torque (i.e. the irreversibility field $\mu_0 H_\mathrm{irr} = \mu_0 H_\mathrm{irr}(\theta) \cos\theta$, where $\theta$ is the angle between the applied magnetic field and the crystalline $c$-axis), up to the highest accessible static magnetic fields of 45 T as expected for bulk superconductivity (Figs. 2$A$, 4A, Materials and Methods, and Figs. S3-S5) (refs. 5–8). Single crystals of the same doping from different batches and grown by different groups show similarly high irreversible magnetic fields in agreement with the high resistive magnetic fields we access in this work (Fig. S3).

We provide further evidence that argues against an origin of magnetic field-resilient super-conductivity in the ground state of underdoped $YBa_2Cu_3O_{6+x}$ from high critical temperature doping inclusions. Superconducting homogeneity is indicated by the narrow superconducting transition in the electrical resistivity as a function of temperature in high magnetic fields (Fig. 1*I,J*). Further support for superconducting homogeneity is provided by the sharpness of the transition in the magnetic susceptibility ($\chi$) in very small magnetic fields (Figs. S6, S7), from which we infer a minimal volume fraction of any regions of the sample with a value of $T_c$ greater than the mean $T_c$ (defined by the step in $\chi$). The observation of low critical temperatures at high magnetic fields further argues against the inclusions of higher dopings as responsible for the persistence of superconductivity up to high magnetic fields. The systematic doping evolution of the high-field superconducting region, which reaches higher critical temperatures with increasing doping also supports the intrinsic bulk character of the high magnetic field–resilient superconductivity (Fig. 2).

In order to discern the nature of the high-field superconducting ground state, we perform a study of the voltage-current characteristics, signatures of which are used to characterize regimes of superconducting vortex physics.[9] We find a striking and systematic non-ohmic voltage-current dependence at high magnetic fields and low temperatures (Fig. 3*A-E*). Given that previous measurements in the vortex state were largely confined to low magnetic fields and high temperatures,[9] we use a model of unconventional quantum vortex matter developed to treat the high magnetic field region to compare with our measurements in the high magnetic field–low temperature region of underdoped $YBa_2Cu_3O_{6+x}$.[10] We find the measured non-ohmic voltage-current dependence can be well captured by a model of quantum vortex matter based on self-organisation of vortices in a magnetic field.[10,11] We use the term 'quantum vortex matter' to describe the vortex regime in high magnetic fields and low temperatures where quantum fluctuations are expected to be relevant,[10] as opposed to the more conventional vortex regime at low magnetic fields and high temperatures.[9] We extract a temperature scale[10] ($T_{HFF}$) as-sociated with the melting of quantum vortex matter into a vortex liquid (Fig. 3*F*) for various magnetic fields, and find that the quantum vortex matter–vortex liquid phase boundary agrees well with the extracted resistive magnetic field $\mu_0 H_r$ below which the voltage drops to a vanishingly small value for a range of temperatures (Fig. 2*A*). Our findings thus unearth a bulk quantum vortex matter ground state that persists up to at least 45 T and evolves to a vortex liquid with increasing temperature as shown in the phase diagrams (Figs. 2*A-C*, 4*E*, SI Appendix Fig. S8). A similar superconducting phase diagram driven by quantum fluctuations has been reported in two-dimensional materials families such as the organic superconductors,[12] and may be similarly expected in the strongly interacting quasi-two dimensional cuprate superconductors. Similarly non-BCS (Bardeen-Cooper-Schrieffer)-like magnetic field-resilient superconductivity with positive curvature of the resistive magnetic field was reported in high purity single crystals of $Tl_2Ba_2CuO_{6+\delta}$.[3] An interplay of superconductivity and a density wave[13] has been further proposed to yield a steep magnetic field – temperature slope of the superconducting phase boundary, as observed in our experiments.

The finite resistivity previously accessed above a modest magnetic field scale $\approx$ 22 T in underdoped $YBa_2Cu_3O_{6+x}$[1] can be attributed to the use in pulsed magnetic field experiments of applied measurement current densities three orders of magnitude higher than present measurements, and even larger eddy current densities (Methods) at elevated temperatures (Fig. S9), yielding vortex dissipation. Previous heat capacity measurements were performed at elevated temperatures and do not access low enough temperatures to capture the unconventional quantum vortex matter regime (Fig. S9, ref. 14). Features in thermal conductivity previously interpreted as a signature of the upper critical magnetic field in $YBa_2Cu_3O_{6+x}$ (ref. 1) meanwhile differ from signatures of the upper critical magnetic field as observed in other type-II superconductors (Fig. S10), prompting its alternative interpretation as a density wave transition in $YBa_2Cu_3O_{6+x}$ (ref. 5).

The superconducting phase diagram in high-magnetic field–low-temperature space for the underdoped cuprates revealed by our present measurements is shown in Fig. 4*E* and Fig. S11. We find the quantum vortex matter region characterized by vanishing electrical resistivity in the $j \rightarrow 0$ limit and non-ohmic voltage-current characteristics (coloured shading) to steeply rise at low temperatures, persisting beyond the highest laboratory accessible static magnetic fields of 45 T. The newly uncovered high magnetic field superconducting phase diagram supercedes previous proposals involving a finite electrical resistivity ground state (Fig. S8*A,B*). Previous proposals include a BCS-like type-II superconducting phase diagram in which a Meissner superconducting state rapidly enters a conventional metallic region[1,14] at high magnetic fields via a vortex solid (Shubnikov phase) region[9] (Fig. S8*A*, S12), and the alternative possibility of a vortex liquid ground state at high magnetic fields characterized by finite electrical resistivity[5,15,16] (Fig. S8*B*).

We gain insight into the character of the quantum vortex matter ground state of the pseudogap by examining the quantum oscillations that are hosted in this region characterized by hysteretic magnetic torque (zero applied measurement current) evidencing vortex pinning,[5,6] vanishing electrical resistivity in the $j \rightarrow 0$ limit, and non-ohmic electrical transport (Fig. 4*A,B*). Quantum oscillations in the electrical resistivity also appear in the quantum vortex matter region, upon the application of sufficiently elevated currents for finite resistivity to be induced from vortex dissipation (Fig. 4*D*). We compare quantum oscillations in the superconducting region of underdoped YBa$_2$Cu$_3$O$_{6+x}$ with those observed in other type-II superconductors including NbSe$_2$, V$_3$Si, Nb$_3$Sn, YNi$_2$B$_2$C, LuNi$_2$B$_2$C, UPdAl$_3$, URu$_2$Si$_2$, CeCoIn$_5$, CeRu$_2$, $\kappa$-(BEDT-TTF)$_2$Cu(NCS)$_2$, MgB$_2$, and others,[17–19] for which theories have been developed of quantum oscillations in the mixed state (e.g. refs. 19, 20). To estimate the extent of superconducting damping of quantum oscillations, we compare the ratio of the Landau level spacing $\hbar\omega_c$ to the maximum superconducting gap $\Delta$ (here $\omega_c = e\mu_0 H/m^*$ is the cyclotron frequency and $m^*$ is the cyclotron effective mass) at magnetic fields where superconducting damping reduces the quantum oscillation amplitude (corrected for the Dingle damping factor) by a factor of two.[21] In the case of underdoped $YBa_2Cu_3O_{6+x}$, we obtain an upper bound for this ratio at the lowest magnetic field value at which quantum oscillations are observed. A low value of $\hbar\omega_c/\Delta \lessapprox 0.06$ is estimated for underdoped $YBa_2Cu_3O_{6+x}$, taking the maximum superconducting gap at zero magnetic field $\Delta \approx 30$ meV from complementary measurements[22] (consistent with the high magnetic field resilience of superconductivity), $m^* = 1.6\ m_e$ ($m_e$ is the free electron mass),[23] and $\mu_0 H = 20$ T. This ratio in underdoped $YBa_2Cu_3O_{6+x}$ is an order of magnitude smaller than the ratio $\hbar\omega_c/\Delta \approx 0.5$ for conventional type-II superconductors including $NbSe_2$, $V_3Si$, $Nb_3Sn$, $YNi_2B_2C$, $LuNi_2B_2C$, $CeRu_2$, $MgB_2$ (see details in Table S1). Quantum oscillations thus persist in the presence of a large maximum superconducting gap, displaying minimal superconducting damping in the case of underdoped $YBa_2Cu_3O_{6+x}$, unlike conventional type-II superconductors. Similarly low ratios of $\hbar\omega_c/\Delta$ as underdoped $YBa_2Cu_3O_{6+x}$ are found in unconventional type-II superconductors such as $URu_2Si_2$ ($\hbar\omega_c/\Delta \approx 0.08$ for $\Delta_{ab}$ and $\approx 0.04$ for $\Delta_c$), $\kappa$-(BEDT-TTF)$_2$Cu(NCS)$_2$ ($\hbar\omega_c/\Delta \approx 0.05$), and $UPd_2Al_3$ ($\hbar\omega_c/\Delta \approx 0.006$ for $\Delta_{ab}$ and $\approx 0.3$ for $\Delta_c$) (see details in Table S1).

Models of quantum oscillations in the presence of a spatially uniform superconducting gap associate a low ratio of $\hbar\omega_c/\Delta$ in unconventional type-II superconductors with an anisotropic d-wave superconducting gap, compared to a higher ratio of $\hbar\omega_c/\Delta$ in the case of conventional type-II superconductors characterized by an isotropic superconducting gap.[24–28] Inspection of quantum oscillations in the quantum vortex matter state of underdoped $YBa_2Cu_3O_{6+x}$, however, reveal distinguishing characteristics that are challenging to reconcile with models of spatially uniform superconductivity. Firstly, in models of spatially uniform superconductivity, whether characterized by isotropic superconducting gapping over the full Fermi surface or d-wave superconducting gapping over the full Fermi surface except at a gapless nodal point, the quantum oscillation amplitude is expected to exhibit a reduced temperature variation at low temperatures due to the vanishing of in-gap states.[9, 19–21, 24, 29] In contrast, the quantum oscillation amplitude increases at low temperatures in underdoped $YBa_2Cu_3O_{6+x}$, consistent with the Lifshitz-Kosevich form[30] even within the quantum vortex matter state, signalling Fermi-Dirac statistics of low energy excitations within the gap (Fig. 4*C*). Secondly, models of spatially uniform superconductivity are expected to yield increased damping both as the system transitions from the 'normal' metallic regime in which the superconducting order parameter is destroyed to the vortex liquid regime in which vortices are mobile, and as the system further transitions to the quantum vortex matter regime in which vortices are collectively pinned (Fig. 4*E*). In our present experiments, we access quantum oscillations as the system transitions from a mobile vortex liquid state to the pinned quantum vortex matter state (Fig. 4*B,E*). In contrast to the expectation from models of spatially uniform superconductivity, no discernible additional damping beyond that in the usual Lifshitz-Kosevich description is observed as the quantum oscillations evolve from the vortex liquid regime to the quantum vortex matter regime (Fig. 4*B,C*, Fig. S14, Fig. S13). These properties of the quantum oscillations we observe in the quantum vortex matter regime reveal the coexistence of finite gapless excitations with a large maximum superconducting gap.

The observed coexistence can potentially be explained by nonuniform models of superconductivity that are spatially modulated at a finite wavevector, such as the pair density wave (PDW) recently reported in experiments such as scanning-tunneling microscopy.[31, 31–36, 36–38, 38–44] Unlike models of spatially uniform superconductivity that are characterised by nodal points,[9, 19–21, 24, 29] models of finite wavevector superconductivity result in 'nesting' over only a portion of the Fermi surface and consequently yield a partially gapped Fermi surface and lines of gapless excitations.[31] PDW models display a nodal–antinodal dichotomy in which the antinodes are gapped by a large maximum superconducting gap, while gapless 'Fermi arcs' occur near the nodes.[31,32,34,37] The reconstruction of the gapless nodal 'Fermi arcs' in PDW models yield a sharply defined nodal Fermi pocket, providing a possible explanation for our observation of quantum oscillations hosted in a quantum vortex matter ground state, which are largely undamped by the large maximum superconducting gap. Alternatively, a nodal Fermi pocket has been modelled to arise from Fermi surface reconstruction by biaxial charge density wave order.[45] Our observations potentially point to quantum oscillations in the quantum vortex matter regime due to the interplay of superconductivity and biaxial charge density wave order.

Any model of quantum oscillations in the unconventional quantum vortex matter ground state of the pseudogap region must also explain features such as the isolated nodal Fermi pocket found by complementary observations of forward–sawtooth form quantum oscillations,[2] a low measured value of linear specific heat capacity in high magnetic fields,[46] the high magnetic field saturation of quantities such as the specific heat capacity[1] and the spin susceptibility from the Knight shift.[47] An open question pertains to the extent to which vortex physics persists over the broader doping, temperature, and magnetic field range of the pseudogap region of the underdoped cuprate phase diagram.[31,34,48]

# Materials and Methods

## Sample preparation for transport measurements

The electrical transport is measured on pristine detwinned oxygen-ordered single crystals of YBa$_2$Cu$_3$O$_{6+x}$ grown by the flux technique. Samples with typical dimensions of (0.8 - 1.5) mm × (0.5 - 1.0) mm × (0.03 - 0.08) mm were selected for the electrical transport measurements. Gold pads of standard six-contact geometry were deposited onto the top surface with 160 nm thickness and to the sides with 80 nm thickness using thermal evaporation methods. Top and side views of a typical transport sample are shown in SI Appendix Fig. S1. Samples with gold pads were annealed at temperatures above 500 °C with flow of high-purity oxygen (> 99.9999 %) to set the oxygen content $x$ and meanwhile allow the gold pads to diffuse into the bulk of the crystal. All measurements in this work were performed with current flowing along the $\hat{a}$-axis, with crystals detwinned under uniaxial stress of 100 MPa at 250 °C. Cu-O chain superstructures were formed in samples under vacuum conditions of below $3 \times 10^{-2}$ mbar. Samples with current contact resistances of ≈ 0.5 Ω, made using gold wires attached with DuPont 4929N, were used for high-field measurements. SI Appendix Fig. S2 shows the superconducting transitions in the susceptibility for the measured samples, with transition widths similar to previous reports. Hole doping $p$ is inferred from the critical temperature $T_c$,[49] defined as the mid-point of the superconducting transition.

## Critical current densities inferred from magnetic torque and resistivity

Assuming the current to be uniformly distributed throughout the sample and to be flowing predominantly within the CuO$_2$ planes, the critical current density is estimated using $j_c \approx 6\Delta M/\pi r$ where $\Delta M$ is the hysteresis of the magnetisation between the up and down field sweeps in units of Am$^{-1}$, $r \approx \sqrt{lw/2}$ is the effective radius of the sample and where $l$, $w$ and $t$ are the length, width and thickness of the sample along the $\hat{a}$-axis, $\hat{b}$-axis and $\hat{c}$-axis, respectively.[50] In the electrical transport measurements, we also assume the current to be uniformly distributed, from which we obtain the current density of $j \approx I/wt$. We find an order of magnitude agreement between the critical current density estimated from torque hysteresis and from electrical transport measurements (Fig. S5).

## Evidence for bulk superconductivity

An important question concerning the observation of superconductivity that persists up to high magnetic fields is whether the superconductivity is of bulk character. Fig. 1*I,J* presents measurements of the electrical resistivity versus temperature for $p = 0.108, 0.132$. The narrow absolute width of the superconducting transition in electrical resistivity at high magnetic fields indicates no significant increase in inhomogeneity of the superconducting state in strong magnetic fields when $T_c$ is suppressed. Furthermore, an inclusion of a small superconducting volume fraction of a higher doping, were it to exist, would manifest in magnetic susceptibility measurements, which is not observed, as shown in SI Appendix Fig. S3.

The observation of significant vortex pinning in the magnetic torque (i.e. hysteresis in Fig. 4*A*, SI Appendix Fig. S4*C*) accompanying vanishing electrical resistivity in the $j \to 0$ limit below the superconducting transition indicates a bulk superconducting state. No such sharp transition or hysteretic signature in the bulk magnetic torque is expected to occur for surface or filamentary superconducting states,[8] which furthermore typically exhibit small finite electrical resistivity rather than the vanishing electrical resistivity that is observed. The above findings, and the systematic nature and sample independence of our results point to the intrinsic, bulk character of the magnetic field-resilient superconducting state in the pseudogap ground state characterised by low critical temperature and low critical current that we find to be persistent beyond the highest accessible static magnetic fields of 45 T.

## Agreement with complementary measurements

In addition to the sensitivity to elevated temperatures, we find the superconducting state characterized by vanishing electrical resistivity in the $j \to 0$ limit to give way to a vortex liquid state with finite but low electrical resistivity upon using larger measuring current densities of $\approx 10$ Acm$^{-2}$ as used in earlier studies of electrical resistivity,[1] and by performing measurements in rapidly changing pulsed magnetic fields as previously reported, that generate large eddy currents in the sample of the order of $\approx 1000$ Acm$^{-2}$. The phase boundary demarcated by resistive magnetic field ($\mu_0 H_r$) above which the vanishing resistivity vortex matter ground state evolves into a finite resistivity vortex liquid as a function of applied measurement current and temperature shows agreement with results of previous measurements (Fig. S9) at high applied measurement currents and high temperatures. Our measurements are in agreement with the observed $\sqrt{\mu_0 H}$ field dependence of the linear specific heat coefficient extending up to the highest accessible static magnetic fields,[46] which is characteristic of a vortex state.[51] Other specific heat measurements at high magnetic fields[14,52] are limited to relatively high temperatures of 2 K, not low enough to capture the steep upturn observed for $\mu_0 H_r$ at low temperatures. Fig. S10 shows the feature in thermal conductivity in underdoped YBa$_2$Cu$_3$O$_{6+x}$ that has been interpreted as a signature of the upper critical magnetic field.[1] A comparison with the thermal conductivity as a function of magnetic field in other type-II superconductors shows differences in characteristic signatures at the upper critical magnetic field that exhibits an upward slope starting from zero magnetic field.[53] These differences have led to the feature in thermal conductivity in underdoped YBa$_2$Cu$_3$O$_{6+x}$ to be interpreted instead as a signature of a density wave onset at high magnetic fields.[5]

## Low temperature growth of quantum oscillation amplitude compared with Lifshitz-Kosevich expansion

The Fermi-Dirac distribution yields a temperature-dependent quantum oscillation amplitude in the Lifshitz-Kosevich (LK) form.[54] This low temperature growth of quantum oscillation amplitude is given by:

$$R_T = \frac{X}{\sinh X},$$

with $X = 2\pi^2 k_B T m^*/e\hbar\mu_0 H$, where $k_B$ is Boltzmann's constant, $T$ is temperature, $m^*$ is the quasiparticle effective mass, $e$ is the electron charge, and $\hbar$ is the reduced Planck constant.[54] For small $T$, a series expansion of the temperature dependence term yields:

$$R_T \approx 1 - \frac{X^2}{6} + O\left(X^4\right),$$

For small $T$, therefore, the quantum oscillation amplitude linearly increases with decreasing $X^2$ approaching the $T \to 0$ limit. The low temperature growth in quantum oscillation amplitude is captured by the relative change of quantum oscillation amplitude at a finite temperature $A(T)$ with respect to the amplitude at the lowest measured temperature $A_0$, given by:

$$1 - \frac{A(T)}{A_0} = \frac{A_0 - A(T)}{A_0}$$
$$= \frac{X^2}{6}.$$

A plot of $(A_0 - A(T))/A_0$ against $X^2$ would therefore yield a straight line with a gradient equal to 1/6 at low temperatures for low energy excitations within the gap. In contrast, in the absence of low energy excitations, gapped quantum oscillation models would yield a much reduced change in amplitude as a function of $X^2$ at low temperatures well below the gap temperature scale.[24] The inset to Fig. 4C shows the growth in quantum oscillation amplitude plotted against $X^2$ with a quasiparticle effective mass $m^*/m_e = 1.676$ (ref. 55). The rapid low temperature growth of the quantum oscillation amplitude yields a linear slope of 0.20(2) at low temperatures, in notable contrast to the expectation of little to no growth in the case of gapped quantum oscillations in the low temperature limit. A full temperature dependence of the quantum oscillation amplitude up to a temperature of 18 K is shown in Fig. S14.

# Author contributions

Y.-T.H., J.P., T.L., M.L.T., B.K. prepared samples; Y.-T.H., M.H., A.J.D., A.J.H., M.K.C., S.V.T., H.L., A.G.E., H.Z., J.W., Z.Z., N.H., S.E.S. performed measurements; Y.-T.H., M.H., A.J.D., A.J.H., H.L., G. G. L., N.H., S.E.S. analyzed data; N.H., S.E.S. wrote the paper with input from all the co-authors.

# Acknowledgements

Y.-T.H., M.H., A.J.D., A.J.H., S.V.T., H.L. and S.E.S. acknowledge support from the Royal Society, the Winton Programme for the Physics of Sustainability, Engineering and Physical Sciences Research Council (EPSRC; studentship and grant numbers EP/R513180/1, EP/M506485/1 and EP/P024947/1), and the European Research Council under the European Unions Seventh Framework Programme (Grant Agreement numbers 337425 and 772891). A portion of magnetic measurements were carried out using the Advanced Materials Characterisation Suite in the University of Cambridge, funded by EPSRC Strategic Equipment Grant EP/M000524/1. S.E.S. acknowledges support from the Leverhulme Trust by way of the award of a Philip Leverhulme Prize. H.Z., J.W. and Z.Z. acknowledge support from the National Key Research and Development Program of China (grant no. 2016YFA0401704). We are grateful for helpful discussions with colleagues including P. W. Anderson, L. Benfatto, J. Blatter, J. C. S. Davis, N. Doiron-Leyraud, M. Eisterer, E. M. Forgan, R. H. Friend, D. Geshkenbein, P. Kim, S. A. Kivelson, M. H. Julien, D. H. Lee, P. A. Lee, T. Maniv, D. R. Nelson, M. R. Norman, N. P. Ong, C. Pepin, M. Randeria, S. Sachdev, J. Schmalian, T. Senthil, L. Taillefer, C. M. Varma, H. H. Wen. A portion of this work was performed at the National High Magnetic Field Laboratory (NHMFL), which is supported by NSF Cooperative Agreement DMR-1157490, the State of Florida, and the Department of Energy (DOE). M. K. C. and N. H. acknowledge support from the DOE Basic Energy Sciences project: 'Science of 100 tesla'. We thank S. A. Kivelson for suggesting the application of the vortex model of Huse et al. to our data. We are grateful for experimental support at NHMFL, Tallahassee from J. Billings, E. S. Choi, B. L. Dalton, D. Freeman, L. J. Gordon, D. E. Graf, M. Hicks, S. A. Maier, T. P. Murphy, J.-H. Park, K. N. Piotrowski, and others. We thank S. Lacher and C.T. Lin for assistance with synthesis of high-quality single crystals.

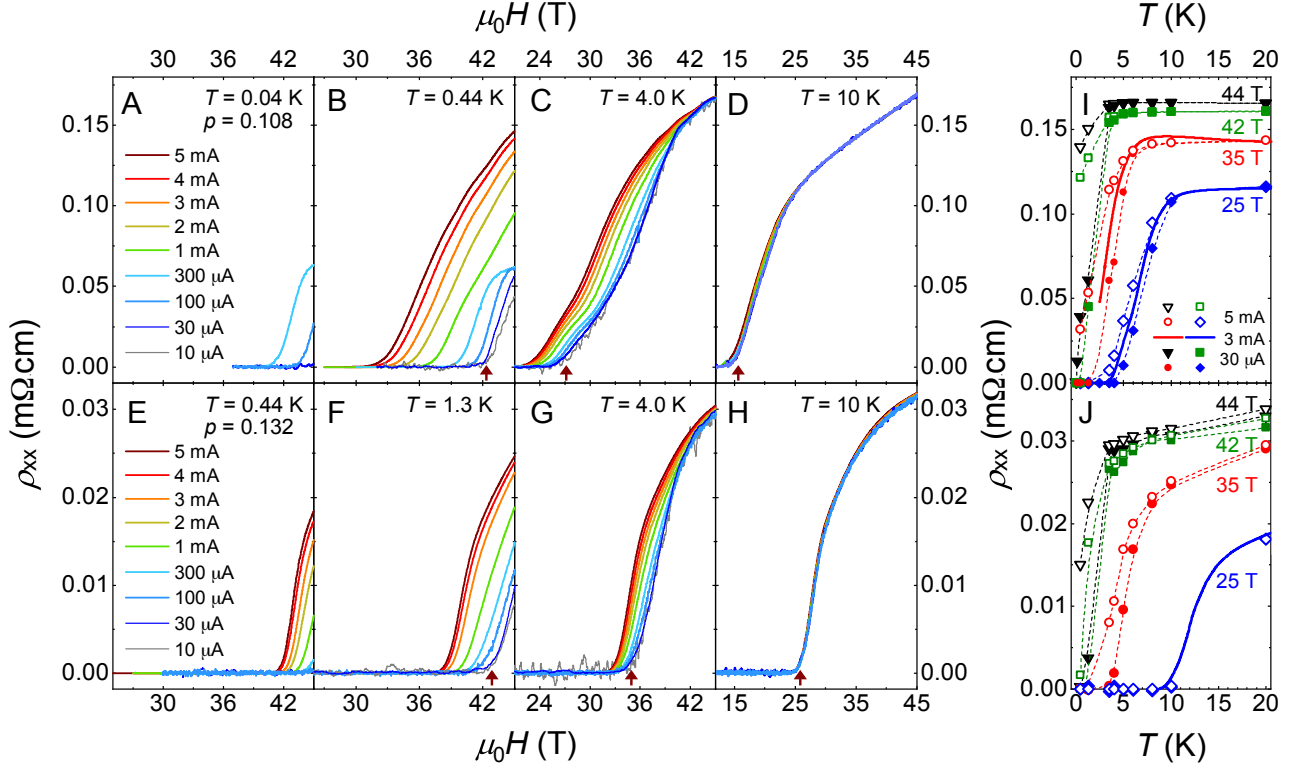

Figure 1: Onset of finite electrical resistivity as a function of magnetic field for different applied measurement currents. (*A-H*) In-plane electrical resistivity $\rho_{xx}$ of YBa$_2$Cu$_3$O$_{6+x}$ measured on sweeping the magnetic field ($\mu_0\mathbf{H} \parallel \hat{c}$) at different temperatures $T$, hole dopings $p$, and applied measurement currents $I$ as indicated. Here $\rho_{xx} = V_{xx}/(jl)$ is an effective resistivity in the non-ohmic regime. Brown arrows indicate the resistive magnetic field ($\mu_0 H_r$) above which the electrical resistivity becomes finite in the limit of low applied measurement current ($\mu_0 H_r(j \to 0)$). $I = 1$ mA corresponds for our samples to a current density of $j \approx 5$ Acm$^{-2}$ on assuming a limit of uniform current distribution. Superconductivity characterized by vanishing electrical resistivity in the $j \to 0$ limit is revealed to persist up to the highest accessible magnetic fields. At the lowest temperature of 40 mK, currents higher than 300 $\mu$A are expected to lead to self-heating effects, and hence only measurements with applied currents below 100 $\mu$A are shown. (*I-J*) In-plane electrical resistivity $\rho_{xx}$ transition into the superconducting state characterized by vanishing electrical resistivity in the $j \to 0$ limit shown at different values of magnetic field (symbols show pulsed field data and dashed lines connecting symbols are guides to the eye, solid lines show DC field data). Width of the superconducting transition is similarly narrow to that observed at zero magnetic fields. Legend in panel *I* also applies to panel *J*.

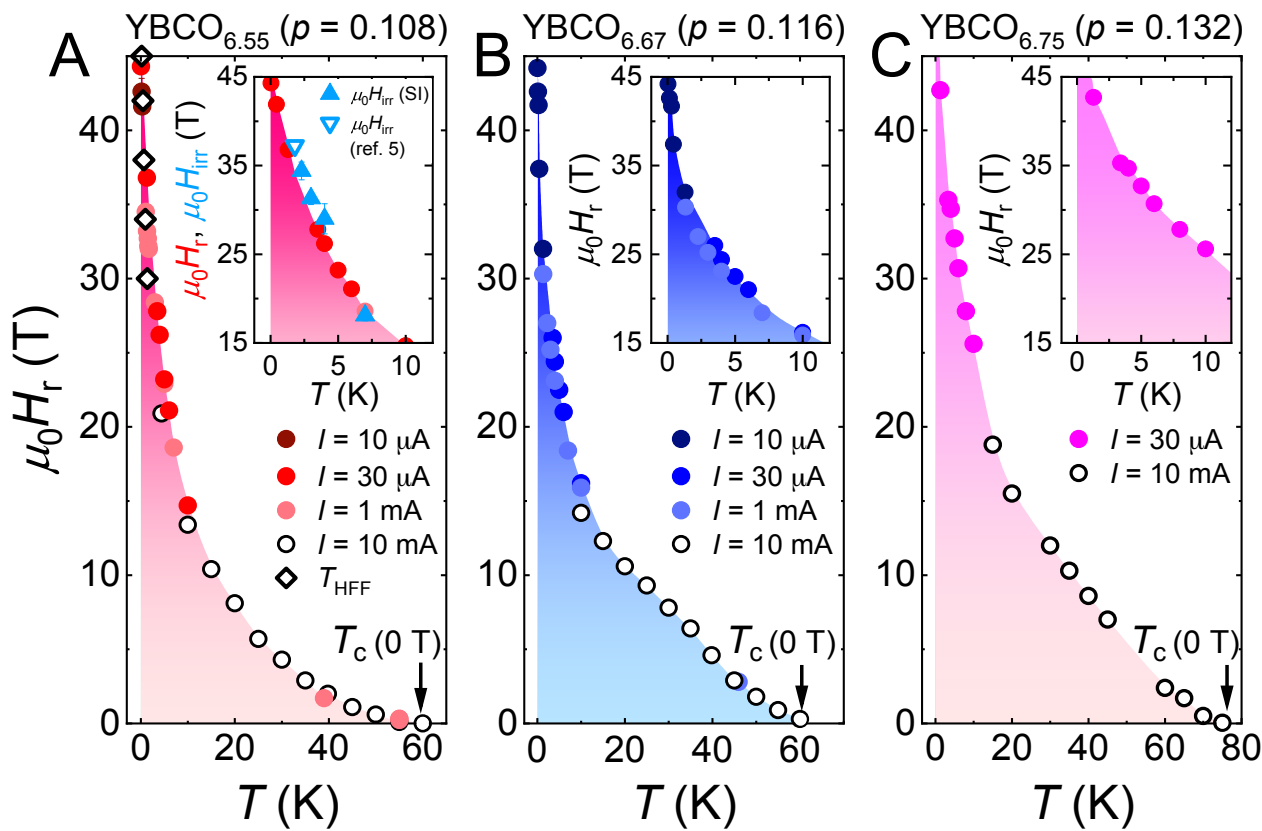

Figure 2: Magnetic field-resilient superconductivity revealed at low temperatures and low applied measurement currents. Resistive magnetic field $\mu_0 H_r$ measured for YBa$_2$Cu$_3$O$_{6+x}$ at three hole dopings ($p$ = 0.108, 0.116, and 0.132) using static magnetic field scans at different fixed temperatures and currents $I$ indicated by circles (here, a current of $I$ = 1 mA corresponds to a current density $j \approx 5$ Acm$^{-2}$), obtained from Fig. 1. Magnetic field-resilient superconductivity is observed to persist up to the highest magnetic fields for low applied measurement currents. Similarly non-saturating values of resistive magnetic field down to the lowest temperatures were reported in Tl$_2$Ba$_2$CuO$_{6+\delta}$ (Fig. S2).[3] The temperature scale associated with the melting of quantum vortex matter into a vortex liquid ($T_{\text{HFF}}$) are indicated by diamonds in panel a. The bulk character of superconductivity at high magnetic fields is indicated by the similar phase boundary obtained from magnetic torque measurements (inset to panel *A*).

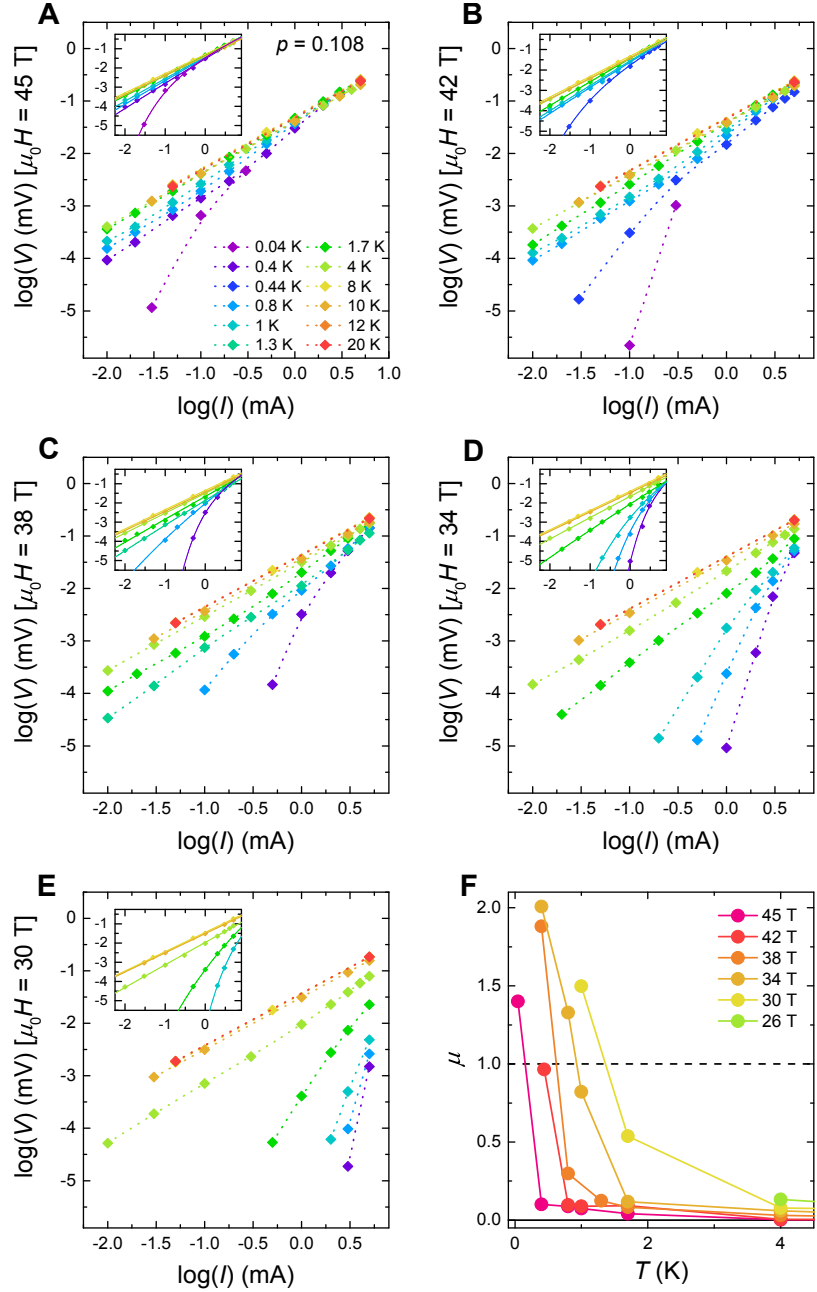

Figure 3: Non-ohmic voltage-current dependence at the highest magnetic fields and lowest temperatures. (A-E) In-plane longitudinal voltage measured for $YBa_2Cu_3O_{6+x}$ ($p$ = 0.108) as a function of applied measurement current, at temperatures ranging from 20 K down to 0.04 K at different applied magnetic fields. Voltage values measured at the same temperature are connected by dashed lines. Solid lines in the insets correspond to fits based on a model of quantum vortex matter by Huse, Fisher, and Fisher,[10] characterized by non-ohmic current-voltage dependence described by $V \propto j \exp[-(j_T/j)^\mu]$, where $j$ is the current density, and $j_T$ and $\mu$ are fitting parameters. (F) The exponent $\mu$ as a function of temperature, with $T_{\text{HFF}}$ defined as the temperature when $\mu = 1$.

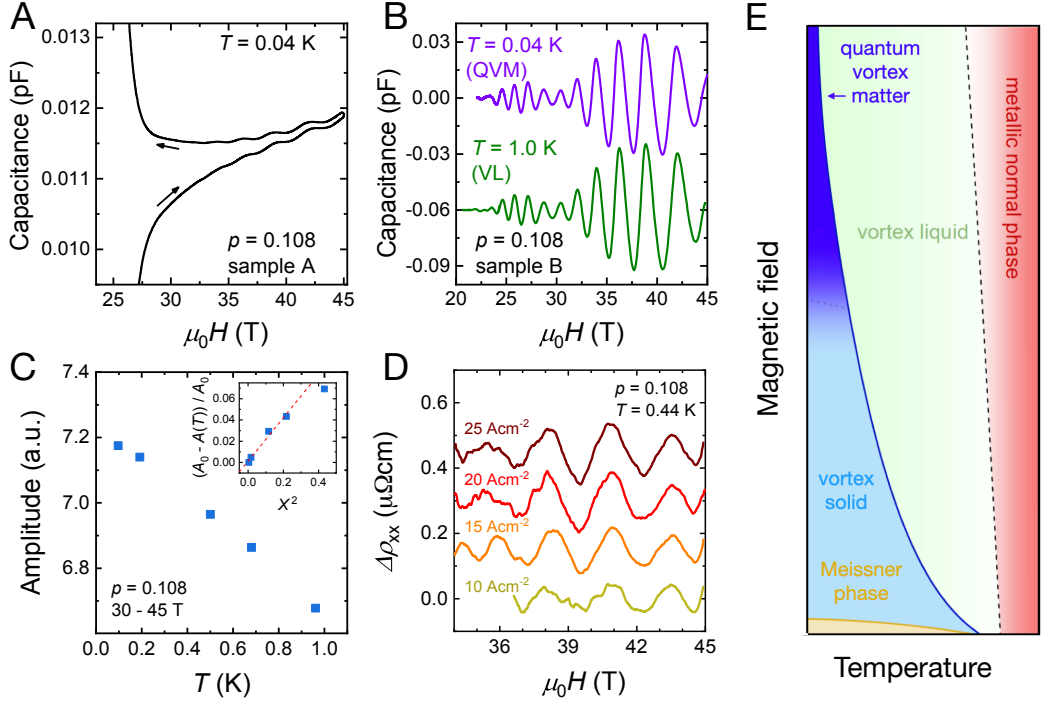

Figure 4: Quantum oscillations co-existing with the vortex matter phase and the magnetic field–temperature phase diagram for $YBa_2Cu_3O_{6+x}$. (*A*) Quantum oscillations co-exist with hysteresis in magnetic torque (zero applied measurement current, $\theta = 9°$) from vortex pinning extending up to the irreversibility field $\mu_0H_{irr}$ beyond 45 T, coincident with the vanishing electrical resistivity region that also extends beyond 45 T at $T = 0.04$ K (also Fig. S3). (*B*) Quantum oscillations in the quantum vortex matter (QVM) regime at the lowest temperature $T = 0.04$ K is seen to be of similar size to quantum oscillations in the finite electrical resistivity vortex liquid (VL) region at elevated temperature $T = 1.0$ K (see also Fig. S13). (*C*) Lifshitz-Kosevich (LK) temperature dependence of the quantum oscillation amplitude at the lowest measured temperatures. The inset shows the growth of quantum oscillation amplitude continues to the lowest measured temperatures, as brought out by a low temperature expansion (Methods). $A(T)$ is the quantum oscillation amplitude at temperature $T$, $A_0$ is the amplitude at the lowest measured temperature, and $X = 2\pi^2 k_B T m^*/e\hbar\mu_0 H$ is the temperature damping coefficient in the LK formula.[54] (*D*) Shubnikov–de Haas oscillations after background subtraction in the quantum vortex matter regime upon applying elevated current densities to induce vortex dissipation. Here $\rho_{xx} = V_{xx}/(jl)$ is an effective resistivity in the non-ohmic regime. (*E*) New superconducting phase diagram in which high magnetic field-resilient quantum vortex matter ground state is revealed in the present measurements, melting to a vortex liquid with elevated temperature.

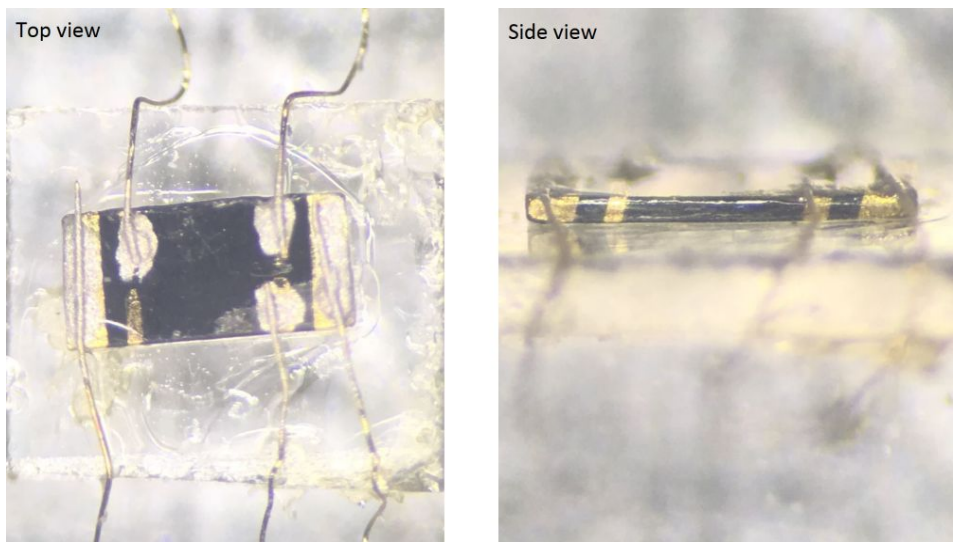

Figure S1: Electrical contacts on measured single crystals of $YBa_2Cu_3O_{6+x}$. Views of a high quality single crystal of $YBa_2Cu_3O_{6+x}$ on which electrical resistivity measurements are performed; quantum oscillations on the same batch of single crystals were reported in ref.[2] Top and side view of electrical contacts attached to the sample mounted on a quartz platelet.

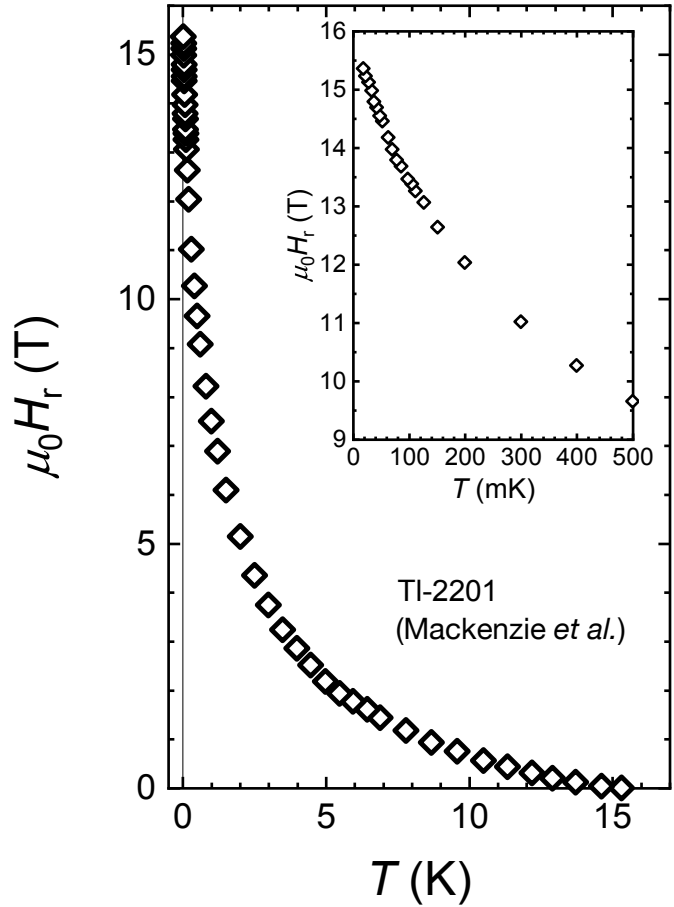

Figure S2: Magnetic field-resilient superconductivity for $Tl_2Ba_2CuO_{6+x}$ (Tl2001, $T_c \approx 16$ K) adapted from ref. 3. It exhibits a sharply rising resistive magnetic field that shows no sign of saturation down to temperatures $\leq 0.001T_c(0)$, similarly to what we report for $YBa_2Cu_3O_{6+x}$. The employed current densities $j$ were from 0.03 to 30 $Acm^{-2}$, with lower current densities employed at the lowest temperatures. Reprinted with permission from ref. 3. Copyright (1993) by the American Physical Society.

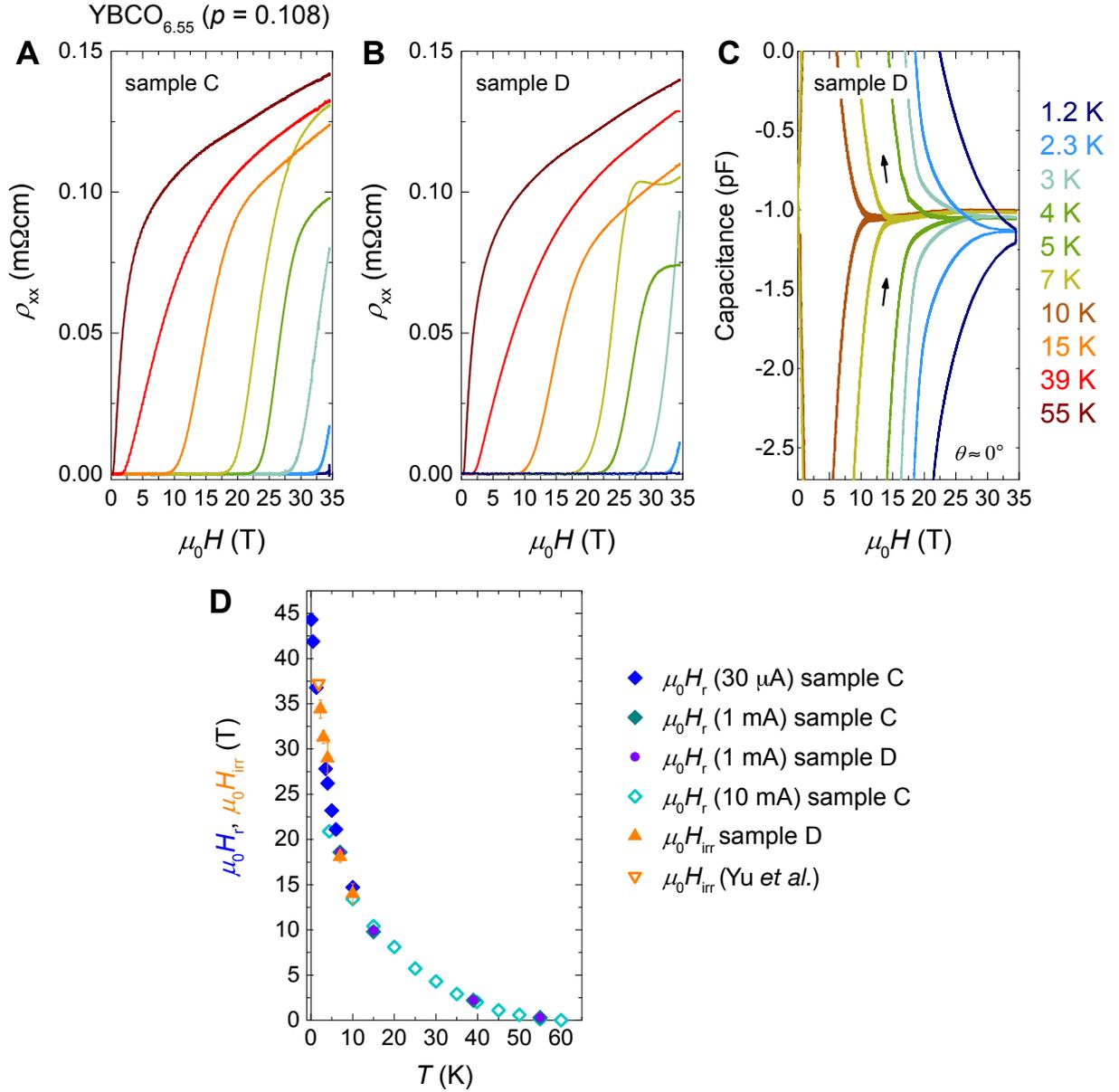

Figure S3: Correspondence of resistive magnetic field ($\mu_0 H_r$) and irreversibility field from magnetic torque hysteresis closure ($\mu_0 H_{irr}$). (A,B) In-plane electrical resistivity $\rho_{xx}$ measured on sweeping the magnetic field, at different temperatures (indicated by colour) using a current of 1 mA for two samples of YBa$_2$Cu$_3$O$_{6+x}$ with $x = 0.55$ ($p = 0.108$). (C) Torque magnetisation measurements of sample D showing hysteretic behaviour between rising and falling fields, revealing the presence of bulk vortex pinning in the same region where the electrical resistivity is vanishingly small, indicative of bulk superconductivity. Here $\rho_{xx} = V_{xx}/(jl)$ is an effective resistivity in the non-ohmic regime. Closure of the hysteresis loop in magnetic torque marks the irreversibility field ($\mu_0 H_{irr}$). (D) $\mu_0 H_r$ obtained from (A,B) and $\mu_0 H_{irr}$ obtained from (C) are seen to coincide for two samples of YBa$_2$Cu$_3$O$_{6+x}$ with $x = 0.55$ ($p = 0.108$). A similar dependence of $\mu_0 H_{irr}$ as a function of temperature up to 45 T was reported in refs. 5, 6. Reprinted from ref. 5.

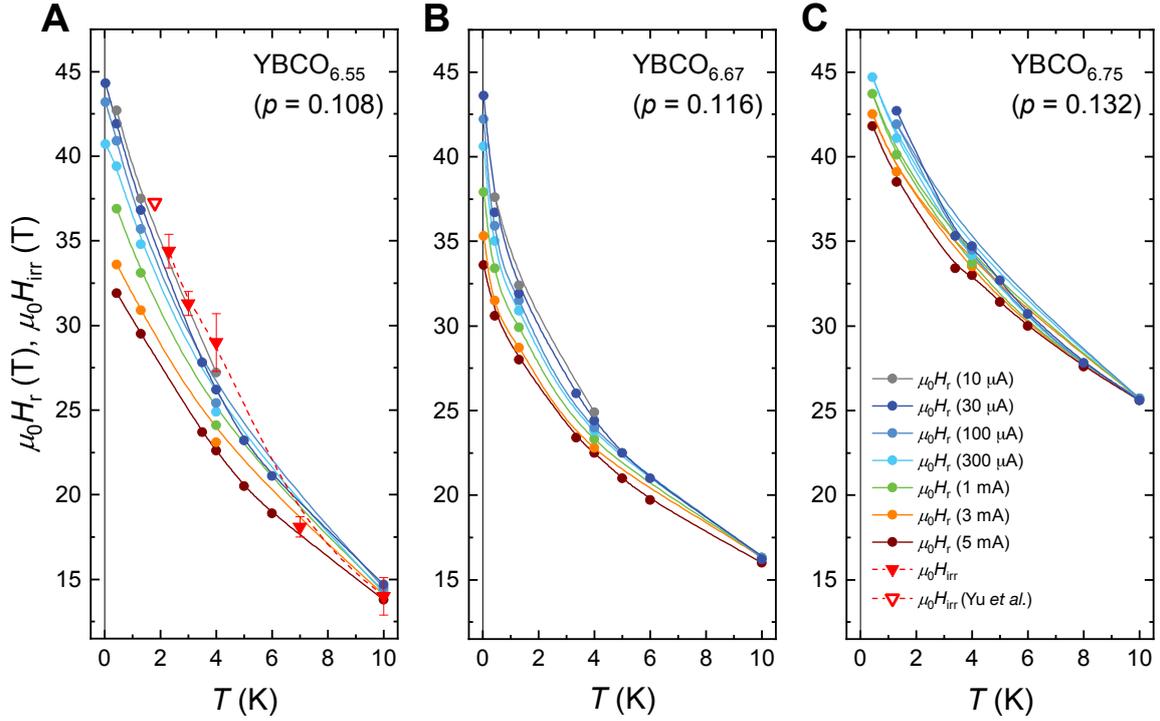

Figure S4: Dependence of resistive magnetic field ($\mu_0 H_r$) on applied measurement current as a function of temperature. (*A-C*) Circles connected by lines indicate $\mu_0 H_r$ observed on measuring the in-plane electrical resistivity as a function of applied static magnetic field at different fixed temperatures and applied currents $I$ for YBa$_2$Cu$_3$O$_{6+x}$ samples with $x$ = 0.55, 0.67 and 0.75. $I$ = 1 mA corresponds to a current density of $j \approx 5$ Acm$^{-2}$. A clear increase in $\mu_0 H_r$ is seen as the applied current is decreased, an effect that grows larger at lower temperatures. Downward triangles show the irreversibility magnetic field measured from hysteretic magnetic torque, reflecting the persistence of bulk pinned vortices up to $\mu_0 H_{irr}$. Good agreement is seen with values of $\mu_0 H_r$ measured from electrical resistivity at the lowest applied current density, confirming the bulk superconducting behaviour reflected by $\mu_0 H_r(j \rightarrow 0)$. Reprinted from ref. 5.

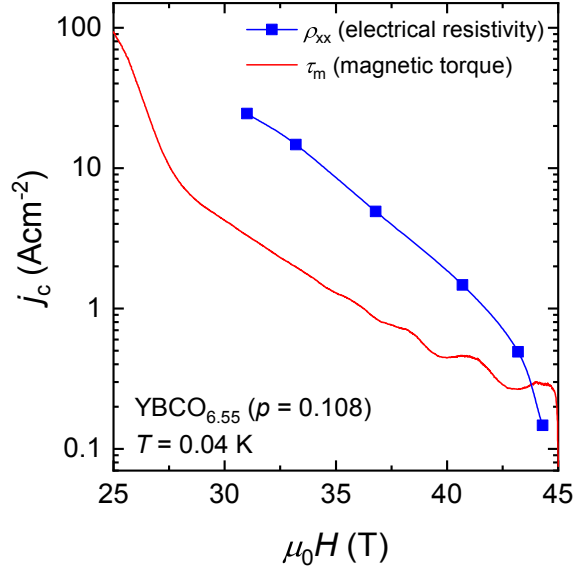

Figure S5: Critical current density inferred from the hysteretic torque magnetisation and resistivity measurements. Fig. 4*A* shows hysteresis between rising and falling field in the measured torque magnetisation for hole doping $p = 0.108$, which is attributed to flux trapped by vortex pinning. We estimate a critical current density from these torque measurements, which is proportional to the extent of hysteresis, using the expression $j_c \approx 6\Delta M/\pi r$ where $\Delta M$ is the hysteresis of the magnetisation between the up and down field sweeps in units of Am$^{-1}$, and $r \approx \sqrt{lw/2}$ is the effective radius of the sample [50] (see Methods). We compare the critical current density obtained from torque magnetisation to the critical current density obtained from transport measurements performed on the same sample, defined as the current density above which the measured resistivity reaches $10^{-3}$ mΩcm, indicating the onset of finite resistivity, and find values obtained are similar to within an order of magnitude, consistent with a bulk superconducting origin.

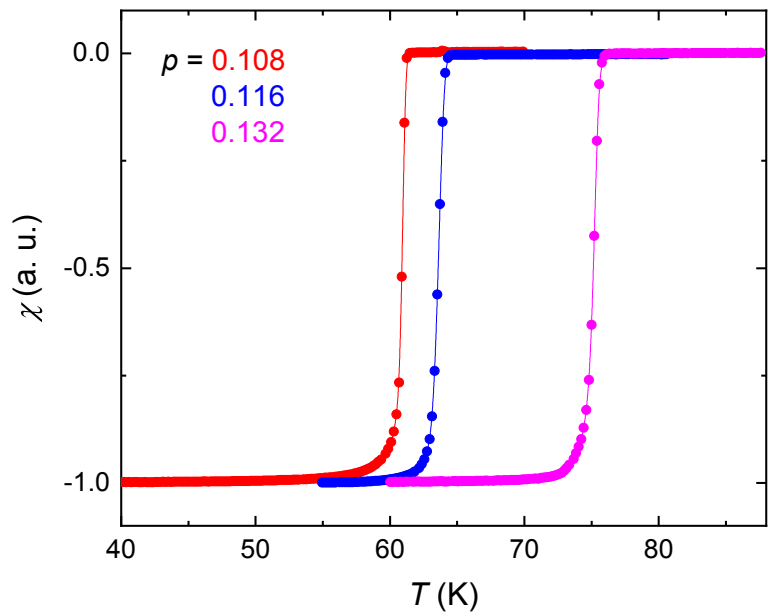

Figure S6: Magnetic susceptibility measurements of the superconducting transition in various dopings of YBa$_2$Cu$_3$O$_{6+x}$. The magnetic susceptibility $\chi$ was measured with a small magnetic field of magnitude 0.2 mT. Sharp superconducting transitions are observed, indicative of high crystal quality and superconducting homogeneity.

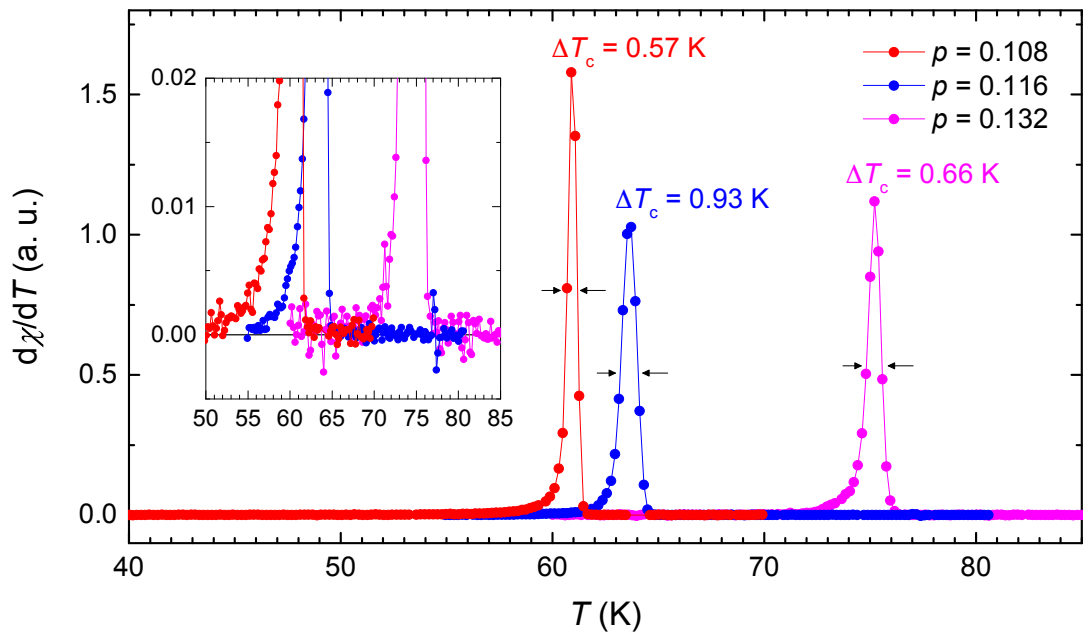

Figure S7: Width of the superconducting transition in underdoped $YBa_2Cu_3O_{6+x}$. The derivative of the magnetic susceptibility $\chi$ is shown for hole dopings 0.108, 0.116, and 0.132 revealing narrow superconducting transitions. The full width at half maximum is found to be less than 1 K for all three dopings.

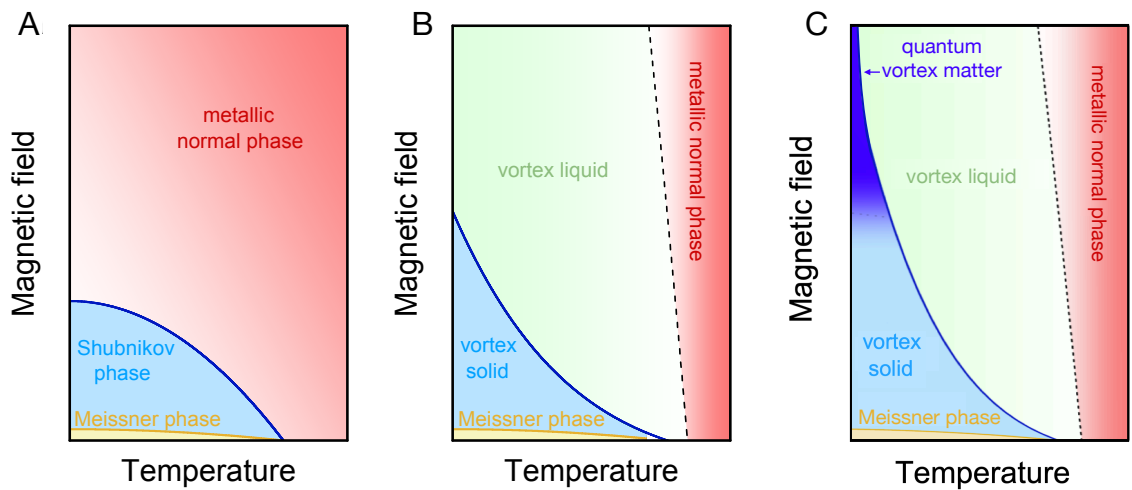

Figure S8: Alternative magnetic field–temperature phase diagrams for $YBa_2Cu_3O_{6+x}$. (*A*) BCS-like type-II superconductor,[1] in which the superconducting order parameter is destroyed above modest magnetic fields, and the normal metallic phase is readily accessed down to the lowest temperatures. (*B*) Proposal for a strongly interacting superconductor from refs.[16,51] in which a vortex liquid phase is accessed above modest magnetic fields. (*C*) New region of high magnetic field-resilient superconductivity of a strongly interacting superconductor is revealed in the present measurements up to the highest accessible magnetic fields, melting to a vortex liquid with elevated temperature.[56,57] Schematic phase diagrams adapted from ref. 9.

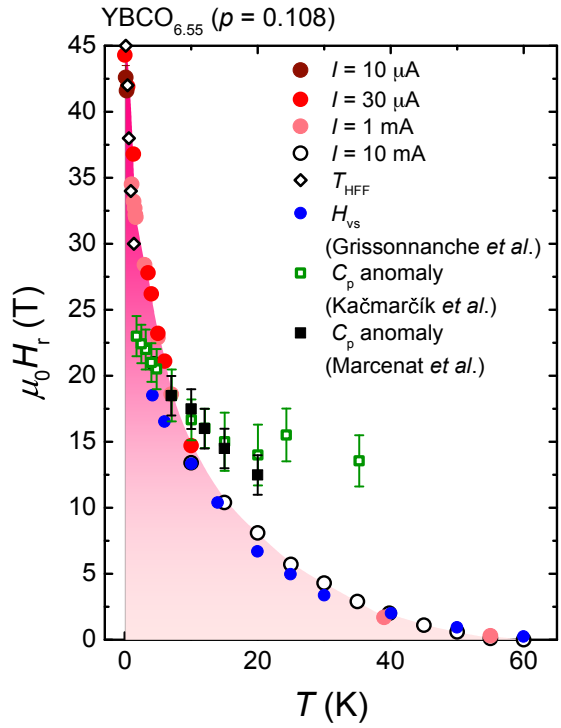

Figure S9: Resistive magnetic field measured in static fields and low applied measurement currents, compared to high current electrical transport measurements and specific heat measurements. The resistive magnetic field $\mu_0 H_r$ measured using static magnetic fields and down to low applied measurement currents revealing magnetic field-resilient superconductivity. The blue circles correspond to $\mu_0 H_r$ obtained from pulsed magnetic field measurements using high applied measurement currents reported in ref. 58. Green and black squares represent the position of the specific heat anomaly inferred from magnetic field scans reported in refs. 14, 52. Specific heat measurements are limited to relatively high temperatures of 2 K, not low enough to capture the steep upturn observed for $\mu_0 H_r$ at low temperatures. Reprinted from ref.52. Copyright (2018) by the American Physical Society. Reprinted from ref.14, which is licensed under CC BY 4.0¡https://creativecommons.org/licenses/by/4.0/legalcode¿. Reprinted from ref.58, which is licensed under CC BY 3.0¡https://creativecommons.org/licenses/by-nc-nd/3.0/legalcode¿.

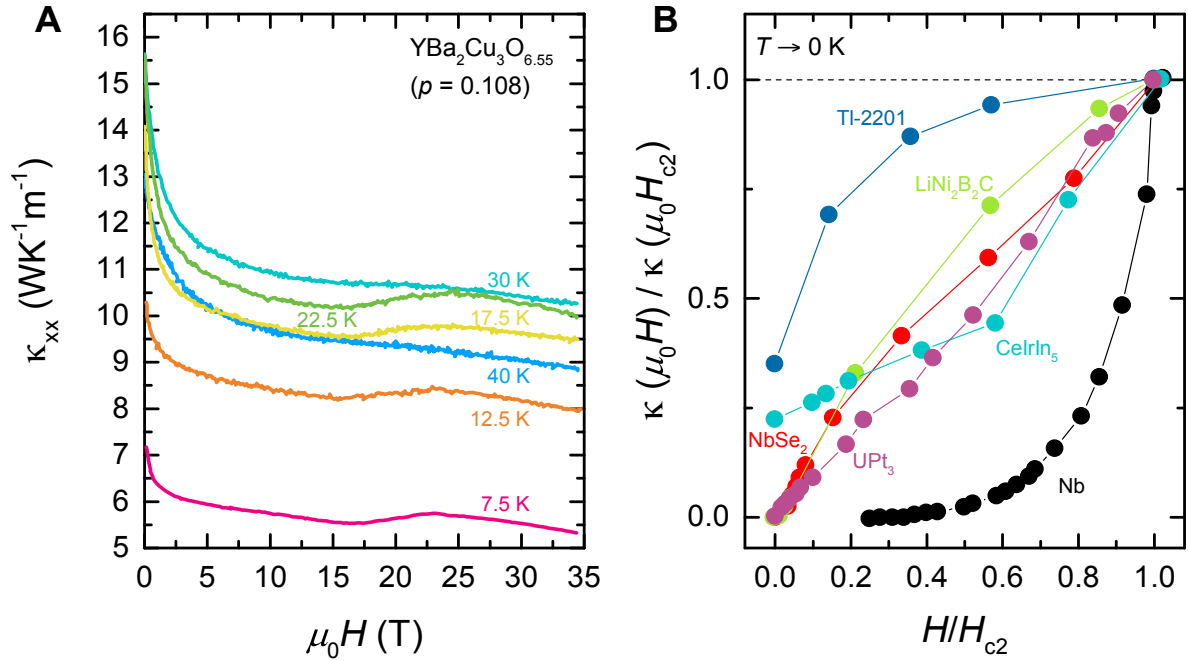

Figure S10: Thermal conductivity as a function of applied magnetic field for $YBa_2Cu_3O_{6.55}$ and type-II superconductors. (*A*) Field dependence of the thermal conductivity of $YBa_2Cu_3O_{6.55}$ at different temperatures, with maximal values at 0 T. Reprinted from ref. 5. It does not show the features observed for type-II superconductors, characterized by a gradual increase in thermal conductivity with applied magnetic field, indicating a transition from the vortex state to the normal state at $\mu_0 H_{c2}$. (*B*) The thermal conductivity of type-II superconductors as a function of applied magnetic field divided by the critical field $\mu_0 H_{c2}$. It shows a gradual increase of the thermal conductivity as a function of applied magnetic field within the vortex region. Reprinted from ref. 53, which is licensed under CC BY 4.0¡https://creativecommons.org/licenses/by/4.0/legalcode¿.

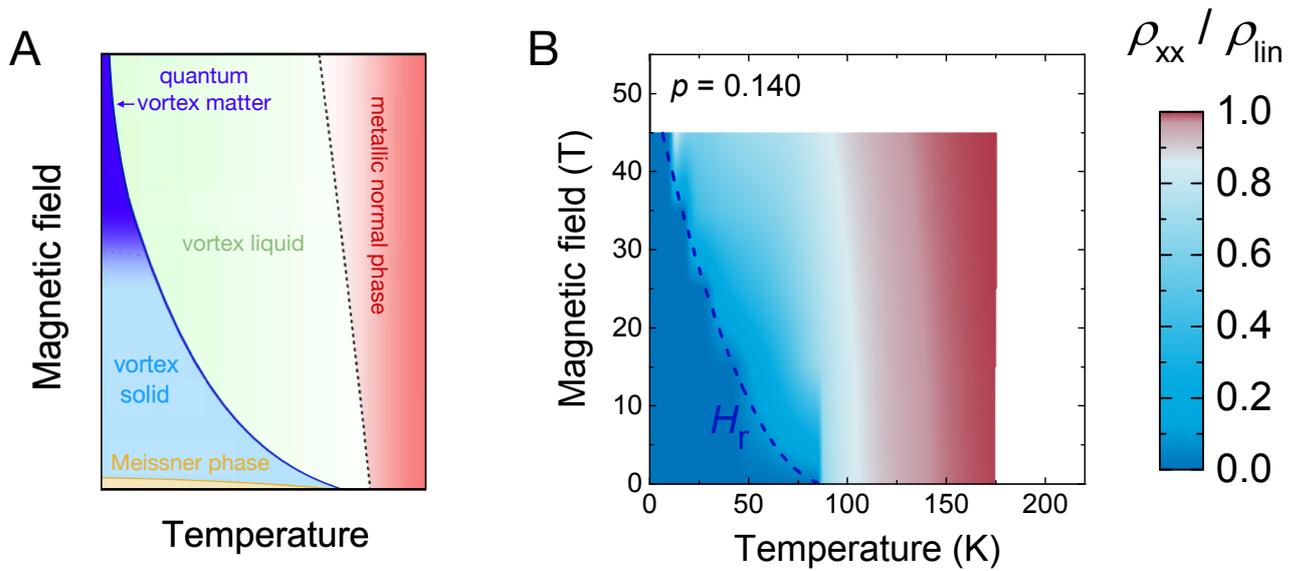

Figure S11: Comparison of the phase diagram of a strongly interacting superconductor with the phase diagram of underdoped $YBa_2Cu_3O_{6+x}$ inferred from resistivity measurements. (*A*) Schematic magnetic field–temperature phase diagram of a strongly interacting superconductor, with a new region of high magnetic field-resilient superconductivity and a vortex liquid region characterized by suppressed resistivity extending over a broad region of magnetic field and temperature. (*B*) Magnetic field–temperature phase diagram constructed from the colour plot of the reduced electrical resistivity of $YBa_2Cu_3O_{6+x}$ hole doping $p = 0.140$ shown in Fig. S12. The resistive magnetic field $\mu_0 H_r$ forms a steep superconducting phase boundary resembling that of the schematic phase diagram in (*A*). We also find an analogous broad region of suppressed resistivity corresponding to the vortex liquid region.

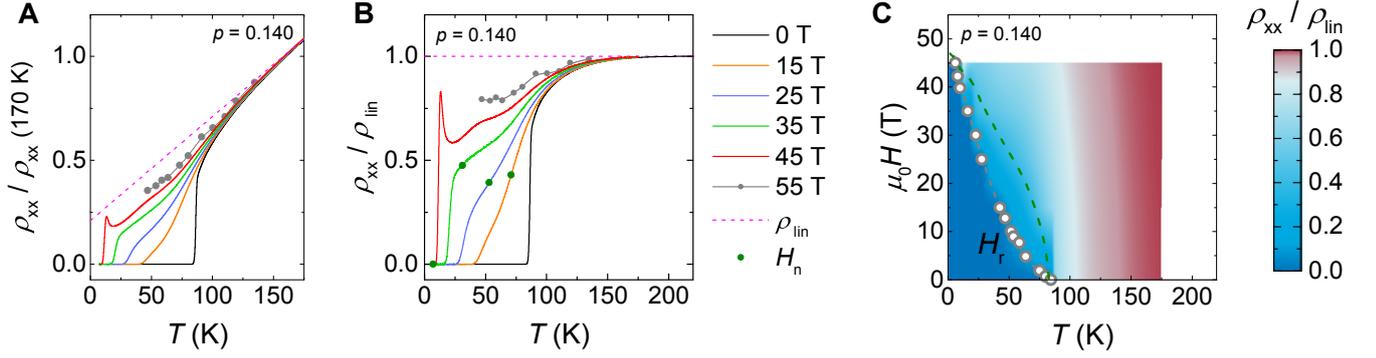

Figure S12: Evolution of the reduced electrical resistivity region up to high temperatures due to strong magnetic fields. (*A*) In-plane resistivity $\rho_{xx}$ of YBa$_2$Cu$_3$O$_{6+x}$ hole doping $p = 0.140$ as a function of temperature for different magnetic fields applied perpendicular to the crystal plane ($\mu_0 H \parallel \hat{c}$). A linear fit ($\rho_{lin} = \rho_0 + a'T$) is made to approximate the temperature dependence of resistivity at high temperatures (pink dashed line). (*B*) Reduced in-plane resistivity from the normal-state resistivity $\rho_{xx}/\rho_{lin}$ as a function of temperature for different magnetic fields. Pink line indicates $\rho_{xx}/\rho_{lin} = 1$. Green circles denote the magnetic field $\mu_0 H_n$ previously interpreted as the onset of the normal state from measurements on the related material YBa$_2$Cu$_4$O$_8$ (identified to have hole doping $p = 0.140$),[58] now seen to clearly lie within the reduced resistivity vortex liquid region. (*C*) Magnetic field–temperature phase diagram constructed from the colour plot of the reduced in-plane electrical resistivity shown in (*B*). Red corresponds to $\rho_{xx}/\rho_{lin} = 1$, while dark blue indicates superconductivity characterized by vanishing electrical resistivity in the $j \to 0$ limit. Open circles indicate the finite electrical resistivity onset magnetic field $\mu_0 H_r$. The intervening region of reduced resistivity appears in light blue. The dashed green line represents the previously interpreted upper boundary of the vortex-liquid phase from ref.,[58] which is now seen to clearly lie within the reduced resistivity vortex liquid region.

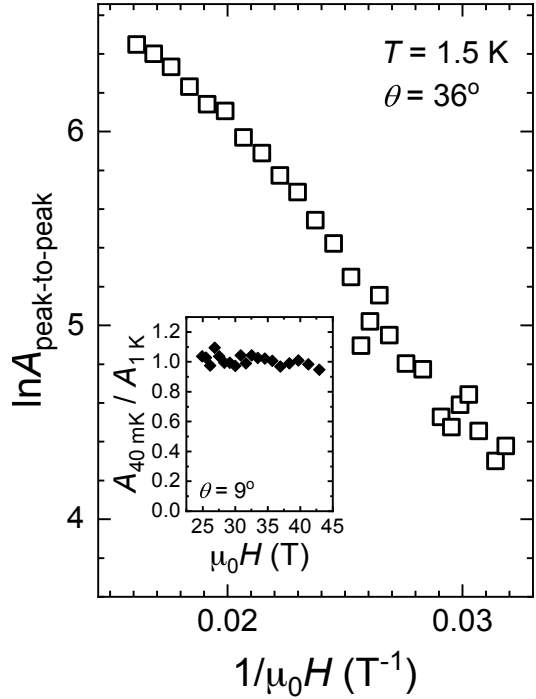

Figure S13: Comparison of quantum oscillation amplitudes in $YBa_2Cu_3O_{6.55}$ in different vortex regimes. Peak-to-peak quantum oscillation amplitude is plotted on a log scale as a function of inverse magnetic field in the vortex liquid regime. Magnetic torque measurements were performed at 1.5 K using the piezocantilever technique in pulsed field up to 65 T, at a tilt angle $\theta = 36°$ off $c$-axis to eliminate beating patterns, on a $YBa_2Cu_3O_{6+55}$ crystal prepared from the same batch as shown in the main text. Inset shows the ratio of quantum oscillation amplitude measured at 0.04 K (quantum vortex matter regime) and 1 K (vortex liquid regime) in this work, as shown in Fig. 4*B*. The quantum oscillation amplitude ratio is found to be close to unity over the entire measured magnetic field range. Reprinted from ref. 2.

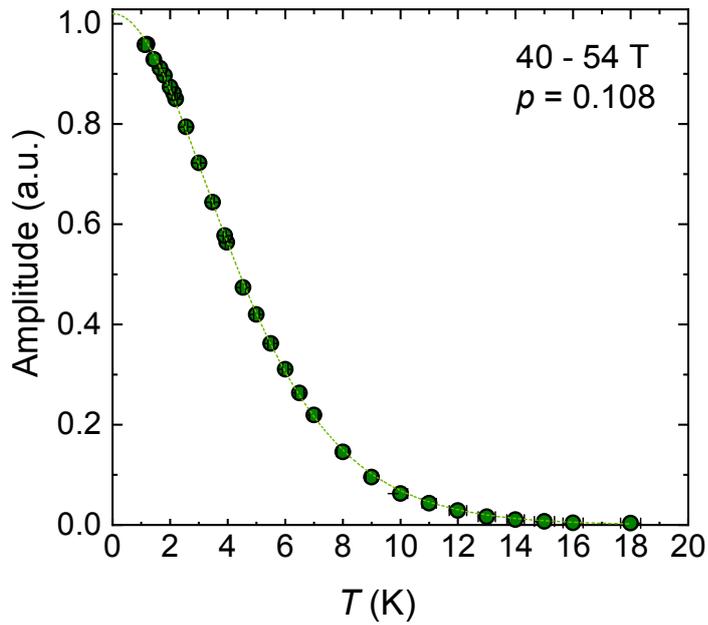

Figure S14: Lifshitz-Kosevich temperature dependence of quantum oscillation amplitude in $YBa_2Cu_3O_{6+x}$ (hole doping $p = 0.108$). Quantum oscillation amplitude measured over a broad temperature range $1.1$ K $\lessapprox T \lessapprox 18$ K, complementing the dilution fridge data below $T \approx 1$ K shown in Fig. 4C; the dashed line shows a fit using a quasiparticle effective mass $m^*/m_e = 1.676$. The quantum oscillation amplitude follows the Lifshitz-Kosevich form down to the lowest temperatures, as expected for quantum oscillations arising from gapless fermionic excitations. Reprinted with permission from ref. 30. Copyright (2010) by the American Physical Society.

Table S1: Parameters characterising superconductors that display quantum oscillations in the superconducting region, including values of the critical temperature $T_c$, magnetic field $\mu_0 H_{50\%}$ where the quantum oscillation amplitude (corrected for the Dingle damping factor) is reduced by a factor of two,[21] the effective mass $m^*$, and the superconducting gap $\Delta$ at zero magnetic field. These values are used to determine the ratio of the Landau level spacing to the superconducting gap.

| Compound | $T_c$ (K) | $\mu_0 H_{50\%}$ (T) | $m^*$ ($m_e$) | $\Delta$ (meV) | $\hbar\omega_c/\Delta$ | Refs. |
|---|---|---|---|---|---|---|
| NbSe$_2$ | 7.2 | 2.3 | 0.61 | 1.2 | 0.4 | 21, 59 |
| V$_3$Si | 17 | 17.2 | 1.3 | 2.6 | 0.6 | 17, 21, 60 |
| Nb$_3$Sn | 18.3 | 14.4 | 1.1 | 3.2 | 0.5 | 17, 60 |
| YNi$_2$B$_2$C | 15.6 | 3.9 | 0.35 | 2.7 | 0.5 | 61, 62 |
| LuNi$_2$B$_2$C | 16.5 | 3.6 | 0.3 | 2.2 | 0.6 | 63, 64 |
| MgB$_2$ | 38.5 | 9.0 | 0.46 | 2.5 | 0.9 | 65 |
| CeRu$_2$ | 6.2 | 3.6 | 0.55 | 1.3 | 0.6 | 66, 67 |
| UPd$_2$Al$_3$ | 2 | 3.6 | 5.7 | 0.24 | 0.3 | 68, 69 |
| URu$_2$Si$_2$ (ab-plane) | 1.5 | 7.2 | 9.5 | 0.7 | 0.13 | 70, 71 |
| URu$_2$Si$_2$ (c-axis) | 1.5 | 2.8 | 13 | 0.3 | 0.08 | 70, 72 |
| $\kappa$-(BEDT-TTF)$_2$Cu(NCS)$_2$ | 10.4 | 4.5 | 3.5 | 3 | 0.05 | 73–75 |
| YBa$_2$Cu$_3$O$_{6.55}$ | 61 | 20* | 1.6 | 20 – 30 | 0.05 – 0.07 | 22, 45, 76–78 |

* The lowest magnetic field value, at which oscillations are observed, giving an upper bound estimate for the Landau level to superconducting gap ratio.